# Best Practices for Administering Attitude and Beliefs Surveys

By Adrian Madsen, Sarah B. McKagan, and Eleanor C. Sayre

Physics faculty care about their students learning physics content. In addition, they usually hope that their students will learn some deeper lessons about thinking critically and scientifically. They hope that as a result of taking a physics class, students will come to appreciate physics as a coherent and logical method of understanding the world, and recognize that they can use reason and experimentation to figure things out about the world. Physics education researchers have created several surveys to assess one important aspect of thinking like a physicist: what students believe that learning physics is all about. In this article, we introduce attitudes and beliefs surveys; and give advice on how to choose, administer, and score them in your classes. This article is a companion to "Best Practices for Administering Concept Inventories"[1], which introduces and answers common questions around concept inventories, which are research-based assessments of physics content topics.

## Introduction to attitude and beliefs surveys

### What is an attitudes and beliefs survey?

Attitudes and beliefs surveys are not about whether students like physics, but about how students perceive the discipline of physics or their particular physics course. These surveys measure students' self-reported beliefs about physics and their physics courses and how closely these beliefs about physics align with experts' beliefs. They ask students questions about how they learn physics, how physics is related to their everyday lives, and how they think about the discipline of physics. For example, students may be asked whether they agree or disagree with statements such as:

- I study physics to learn knowledge that will be useful in my life outside of school
- A significant problem in learning physics is being able to memorize all the information I need to know.

Examples of these types of surveys include[2]:
- Colorado Learning Attitudes about Science Survey (CLASS)[3]
- Maryland Physics Expectations Survey (MPEX)[4]
- Epistemological Beliefs About Physics Survey (EBAPS)[4,5]
- Views About Science Survey (VASS)[6,7]
- Colorado Learning about Science Survey for Experimental Physics (E-CLASS)[8]

These surveys ask students to rank statements using a 5-point Likert scale from strongly agree to strongly disagree. The most common way to score these surveys is to collapse students' responses into two categories depending on whether they are the same as an expert physicist would give (called "percent expertlike response" or "percent favorable response"). They are usually given as a pre- and post-test to measure the "shift," or change in students' beliefs over the course of the term. These surveys are intended to measure the how different teaching practices impact students' beliefs by looking at, on average, the change in these beliefs during in the course. They are not meant to assess individual students or to be used as placement exams to place students into physics courses.

## Why should I use an attitude and beliefs survey?

You should use a beliefs survey to assess the shifts in your students' attitudes and beliefs about physics as a result of your course. If your shifts are negative or zero (or weakly positive), you can make changes to your teaching to help develop your students' beliefs to be more expertlike. Research has shown that courses with an explicit focus on modeling or developing students' expertlike beliefs have significantly greater positive shifts in attitudes and beliefs scores (on the CLASS and MPEX) than courses with some focus on developing expertlike beliefs or ordinary courses[9]. Further, standard physics courses often have a negative impact on students' attitudes and beliefs, where after taking a physics course students' beliefs are less expertlike.

Improving your shifts on attitudes and beliefs surveys is not the goal in and of itself. These survey results are a measurement of how your students think about the discipline of physics, and learning physics. What students' believe about physics can influence the way that they approach learning in their physics course[10]. For example, if students believe that learning physics is about memorizing formulas, they will approach it much differently than if they believe that learning physics is about understanding how you model the physical world to understand it and make predictions about it. So, improving your students' attitudes and beliefs about physics can actually help them more successfully learn physics content and help develop their ability to think like a physicist.

## What goes into the development of an attitudes and beliefs survey?

Research-validated attitudes and beliefs surveys have gone through a thorough development and testing process to make sure the survey is valid, reliable, the wording of the questions makes sense to students, and the content being surveyed is valuable according to experts. This process usually involves the following steps:
1. Develop questions based on taxonomy of topics to be surveyed, experts' views of important topics, and/or questions from previous surveys.
2. Test questions in think-aloud interviews with students. Revise questions.
3. Test written version with large numbers of students. Gather expert responses to survey questions to determine expertlike answers.

4. Complete appropriate statistical analyses (checking reliability, establishing groups of questions using factor analysis etc.). Revise again.

In the section below titled, "How do I evaluate how good an attitudes and beliefs survey is?" we discuss a system for determining the level of research validation for attitudes and beliefs surveys, as used on PhysPort.

### How are attitudes and beliefs surveys different than conceptual multiple-choice tests?

Conceptual multiple-choice tests of physics content usually only have one correct answer, and students get a point for getting it correct, and no points for answering incorrectly. On attitudes and beliefs surveys, there is no "correct" answer, per se, but instead an expertlike (also called favorable) answer, which is determined by asking a group of physics experts to answer the survey questions, and choosing the expertlike answer to be the one that most (but not necessarily all) experts agree on. For example, if experts disagree with a statement that physics is about memorizing information, then students who also disagree may earn one point, while students who agree with that statement do not. The overall score on an attitude and beliefs survey is a measure of how much students agree with physicists, whereas the overall score on a conceptual multiple-choice test is a measure of how much physics content students understand. On attitudes and beliefs surveys, to track changes over a term, we look at "shifts" in students' scores, which is simply the difference between the pre- and post-test scores. On conceptual multiple-choice surveys, to track changes over time, it is common to look at normalized gain, which looks at the amount learned divided by the amount they could have learned[1].

### Limitations of attitudes and beliefs surveys

Attitudes and beliefs surveys don't measure how much students like their physics course. Because this is self-report data, we can't know how well the beliefs students report correspond to the ways they actually think about physics. For example, a student might say and really believe "When I am solving a physics problem, I try to decide what would be a reasonable value for the answer," but not do that in real life. Alternately, a student might agree with "A significant problem in learning physics is being able to memorize all the information I need to know." because they are thinking about the structure of the course they are enrolled in, not the practice of learning physics more broadly.

# Choosing an attitudes and beliefs survey

## Which attitudes and beliefs survey should I use?

Pick the survey that best matches the aspects of your students' beliefs and attitudes that you want to know about. In the case that you find several surveys focus on the same aspect of students' beliefs, choose the survey with the strongest research validation and the survey that is most popular among physics faculty in general, and/or those in your departments, so that you can compare your results to others.

There are four assessments about students' beliefs about learning physics in general: The Colorado Learning Attitudes about Science Survey[3] (CLASS), Maryland Physics Expectations Survey[4] (MPEX), Epistemological Beliefs Assessment for Physical Sciences[4,5] (EBAPS), and the Views About Science Survey[6,7] (VASS). There are five assessments about students' beliefs about specific aspects of physics or their own learning, e.g., labs, problem solving etc. These are the Colorado Learning about Science Survey for Experimental Physics[8] (E-CLASS), the Attitudes and Approaches to Problem Solving[11,12] (AAPS), Attitudes about Problem Solving Survey[13] (APSS), the Physics Goal Orientation Survey[14] (PGOS) and the Self Assessment of Learning Gains[15] (SALG). See our Resource Letter (RBAI-2)[2] for a full discussion and comparison of these different attitudes and beliefs surveys. You can get more details about each of these on PhysPort.org/assessments.

## How do I evaluate how good an attitudes and beliefs survey is?

On PhysPort, we help you determine the level of research validation for an assessment based on how many of our research validation categories apply to the RBAI. These categories include:

- Questions based on research into student thinking
- Studied with student interviews
- Studied with expert review
- Appropriate use of statistical analysis
- Administered at multiple institutions
- Research published by someone other than developers
- At least one peer-reviewed publication

RBAIs will have a gold level validation when all seven categories apply. These have been rigorously developed and recognized by a wider research community. Silver-level RBAIs have 5-6 of these categories which apply, and are also well-validated, but are missing 1 or 2 categories of research validation. In many cases, silver RBAIs have been validated by the developers but not the larger community, often because these assessments are new. Bronze-level assessments have 3-4 categories that apply, and are those where developers have done some validation but are missing substantial elements. Finally, research-based validation means that an assessment is likely still in the early stages, so only 1-2 categories apply.

## Administering attitude and beliefs surveys

### How do I administer an attitude and beliefs?

*General Best Practices for Administering Attitude and Beliefs Surveys*

- When talking to your students about an attitudes and beliefs survey, stress that this assessment is designed help you, their instructor, learn about how your course influences the way they think about physics. It is not meant to evaluate individual students; you are looking what the class as a whole believes about learning physics. Also emphasize that there is no right answer. They should answer based on what they personally believe.
- Make it clear that their results will not influence their course grade. Let them know you would like to know about how they think about these questions so they should answer the questions according to what they really believe, not what they think you want them to say.
- Give students the recommended time to take the survey. You can find this on the PhysPort by searching for your specific assessment on the "Assessment" tab[16].
- If you want to make comparisons with other classes the most meaningful, give the survey in its entirety and with the original wording and question order.

*In-class or out-of-class*

You can give the attitude and beliefs surveys in-class or outside of class. Because this is a survey of students' beliefs about physics, maintaining the security of the survey is not important (unlike concept inventories which measure physics content). Giving the survey in class usually results in a higher completion rate than if students take the survey outside of class.

*Online or paper-and-pencil?*

Taking the survey online outside of class doesn't require class time, and the electronic format means that results can be automatically tabulated, saving valuable time. Paper and pencil versions of the survey require more time to tabulate results, but this time can be decreased by the use of scantrons or other similar tabulation systems. You can also upload a spreadsheet of survey results to the PhysPort Data Explorer[17] where they will be automatically analyzed and visualized. Some surveys are available to be taken online[18] or can be entered into your course management system (e.g. Blackboard, Canvas); the details of this administration vary by survey.

*When should I give the survey?*

Give the survey as a pre-test in the first week of the term and again as a post-test in the last week of the term.

*Best Practices for Administering Concept Inventories Online*

**How?**

1. Announce the survey both in class and by email
2. Give a short (three to seven day) time period for taking the survey
3. Include a timer for surveys administered online. If students take substantially less time to complete the survey than their peers, discard their answers.
4. Provide a follow-up email reminder to students who still need to take the survey.

For the CLASS, it is typical to get a 90% pre-test response rate and 85% post-test response rate online.[3]

*Best Practices for Administering Belief Surveys In class, with paper and pencil*

**Where?** Give the survey during any course meeting, e.g., lecture, recitation or lab.

**How?** To make grading more efficient, use scantrons or another similar tabulation system. Especially if you have multiple sections or plan to give the same survey again in future years, print the survey and the answer page separately. That way you can reuse the survey from section to section.

## What incentives should I give my students for taking the survey?

Some instructors have found that giving a small amount of course credit for completing the survey helps increase the completion rate of the survey. You could also choose not to give any credit for completing the survey. In this case, the way you talk about the survey can help motivate your students to complete it.

## What about test security?

Attitudes and beliefs surveys are asking students to self-report their own perceptions of physics. You do not have to worry about students using outside resources or working together with other students outside of class since there is no right answer on these kinds of surveys. Because of this, you do not need to worry about test security with these kinds of surveys.

# Scoring and interpreting my results

## How do I interpret my scores?

*Overall percent favorable/expertlike versus unfavorable/novicelike scores*

Scores are usually calculated as the percentage of questions that students answer as the same way as an expert in physics (percent favorable or expertlike) or the percentage that they answer in the opposite way as an expert physicists (percent unfavorable or novicelike). The percent favorable and unfavorable scores don't always add up to 100% because students can pick "neutral" as an answer choice, which counts as neither favorable or unfavorable. Both percent favorable and unfavorable give you complimentary information. It is much more common to use percent favorable scores to gauge students' beliefs about learning physics.

There are some attitudes and beliefs surveys that are not scored in this standard way (by calculating shifts in percent favorable). These are the EBAPS, VASS, ECLASS, EBAPS, APSS, PGOS, SALG. For more details about scoring for each of these, search for that assessment on PhysPort.org/assessment.

*Shift in favorable/expertlike scores from pre- to post-test*

The "shift" in percent favorable responses is calculated by subtracting the pre-test class average percent favorable from the post-test class average percent favorable. This metric tells you how students' favorable beliefs about physics changed from the start to end of their physics course. We hope that this shift would be positive, indicating students' beliefs improved over the course of theirs physics class, though in many traditional and even reformed intro physics classes, there is a negative shift in percent favorable scores.

Many instructors who teach the same course several times find it extremely useful to document changes they made to their teaching and compare the shifts in students' beliefs over time to determine how the changes they made influenced their students' beliefs about physics. You can also look at specific categories of questions to learn about how changes in teaching influence these different aspects of attitudes and beliefs about physics. Additionally, you can learn about how your students' beliefs vary based on characteristics of the students like major, gender, year in school etc. and how changes to your teaching influence these sub-groups in your class.

While it can be useful to look at differences in attitudes and beliefs based on student demographics in order to make changes to your teaching to improve students' attitudes and beliefs, we caution against overgeneralizing your results. Making broad-brush generalizations about students because of their race or gender based on differences found on your beliefs survey is inappropriate. That said, Adams et al.[3], found that on the Colorado Learning Attitudes about Science Survey, women have less expert-like scores on the statements in the "real world connections," "personal interest," "problem solving confidence," and "problem solving sophistication" categories. Women scored more expert-like on statements in the "sense-making/effort" category. Differences in scores on the other attitudes and beliefs surveys have not been studied. Knowing that your course and/or broader educational environment is unevenly supporting students in developing expert-like beliefs allows you to look for ways you can support students more equitably in developing their attitudes and beliefs.

*Effect size of shift from pre- to post-test*

To get a sense of how important the difference between your pre- and post-test scores are, you can calculate the effect size of the change.  A large effect sizes mean the difference is important; small effect sizes mean the difference is unimportant. It normalizes the average raw shift in a population by the standard deviation in individuals' raw scores, giving you a measure of how substantially the pre- and post-test scores differ.  There are several different measures for effect size depending on the size of your class.  For more information on calculating effect size, see our expert recommendation about it on PhysPort[19].

### Are students really reporting their own beliefs, and not what they think I want them to say?

Most likely, yes. Gray et al.,[20] gave students the CLASS and asked them to answer two questions for each statement, "What would a physicist say?" and "What do YOU think?" They found a rather large difference between students' personal answer and the answer they believe a physicist would give, with students' personal answer being less expertlike. The "personal" scores were similar to CLASS results from the same courses when the CLASS was given with the standard single answer format. So, this study suggests that when students complete the standard single answer format of the CLASS, they are answering based on their own personal beliefs, although they do know what a physicist would say. The other attitudes and beliefs surveys have not been studied in this way.

### How can I quickly and easily analyze my results using the PhysPort Data Explorer?

PhysPort offers a powerful and secure assessment Data Explorer[17] that makes analyzing and interpreting your attitude and beliefs survey results quick and easy. This online tool allows you to securely upload your students' results and visualize them in a variety of ways. You can use the Data Explorer to calculate the percent favorable/unfavorable scores, and shifts in your scores quickly and compare your own assessment results over time as you make changes to your course. You can use the Data Explorer to look at how your students performed on individual questions or clusters[21] of post-test questions to get a coarse-grained sense of the categories where your students had more expertlike attitudes and beliefs. You can then reflect on the way you supported these expertlike attitudes and look for ways to improve your teaching.

Additionally, you can download a report of your results and comparisons that you can use to talk to your colleagues about your course, include in tenure documents, accreditation reports, etc. The system has robust security measures to ensure that your students' assessment data and your and your students' identities are protected. To use the data explorer go to www.physport.org/dataexplorer.